\documentclass[11pt,english]{iopart}
\usepackage[english]{babel}
\usepackage{amsfonts,amssymb,bm}

\usepackage[T1]{fontenc}
\usepackage[latin3]{inputenc}

\newtheorem{theorem}{Theorem}
\newtheorem{proposition}{Proposition}

\newcounter{llista}

\begin{document}
\jl{6}
\title{On the degrees of freedom of a semi-Riemannian metric}
\author{J Llosa\dag\footnote[3]{Postal address: Martí i Franquès, 1; E-08028  Barcelona (Spain)}  and D Soler\ddag}
\address{\dag\ Departament de F\'{\i}sica Fonamental, Universitat de Barcelona}
\address{\ddag\ Oinarritzko Zientziak Saila, Goi Eskola Politeknikoa,			
Mondragon Unibersitatea}
\address{\ Laboratori de F\'{\i}sica Matem\`atica, SCF (Institut d'Estudis Catalans)}
\ead{pitu.llosa@ub.edu}
\begin{abstract}
A semi-Riemannian metric in a $n$-manifold has $n(n-1)/2$ degrees of freedom, i. e., as many as the number of components of a differential 2-form. We prove that any semi-Riemannian metric can be obtained as a deformation of a constant curvature metric, this deformation being parametrized by a 2-form.
\end{abstract}

\section{Introduction \label{S1}}
It is known, since an old result by Riemann \cite{Riem}, that an $n$-dimensional metric has $f=n(n-1)/2$ degrees of freedom, i. e. it is locally equivalent to the giving of $f$ functions. This feature seems to be a non-covariant property, as it is related to
some particular choice of either a local chart or a local base.

A two-dimensional metric  has \,$f = 1$\, degrees of freedom. In this case, however, a stronger, well known result holds \cite{Eissen}, namely, {\it any two-dimensional metric \,$g$\, is locally conformally flat, \,$g = \phi\,\eta  $\,, \,$\phi$\, being the {\rm conformal} deformation factor and \,$\eta$\, the flat metric} (Gauss theorem). In the two-dimensional case, the above Riemann result is {\em intrinsic} and {\em covariant}, i.e., only tensor quantities are involved, and the sole degree of freedom is represented by a {\em scalar}, conformal deformation factor, $\phi$, which only depends on the metric \,$g$.

The question thus arises of, whether or not, for \,$n>2$\, they exist similar intrinsic and covariant local relations between an arbitrary metric \,$g$\,, on the one hand, and the corresponding flat metric \,$\eta $\, together with a set of  \,$f$\, covariant
quantities on the other. 

There is a number of results concerning the diagonalization of any three-dimensional metric: \cite{Diag1}, \cite{Diag2}, \cite{Diag3}, \cite{Diag4} and \cite{Diag5} to quote some of them. However valuable they are, they are not covariant because, in addition to \,$f = 3$\, scalars, an orthogonal triad or a specific coordinate system is {\em also} involved. To our knowledge, the first published result of this kind for $n=3$ is \cite{CllS01}, where the following theorem was proved:
\begin{theorem} 
Any three-dimensional Riemannian metric \,$g$\, may be locally obtained from a constant curvature metric \,$\eta$\, by a deformation of the form 
\begin{equation}  \label{e0}
 g = a \eta + \epsilon\,\mbox{\boldmath$s$}\otimes\mbox{\boldmath$s$} \: \: ,
\end{equation} 
where $a$\, and \,$\mbox{\boldmath$s$}$\, are respectively a scalar function and a differential 1-form. The sign $ \epsilon = \pm 1$, the curvature of $\eta$ and a relationship $\Psi(a,\mbox{\boldmath$s$})$\, between the scalar \,$a$\, and the
Riemannian norm \,$|\mbox{\boldmath$s$}|$\, may be arbitrarily prescribed.
\end{theorem}

After realizing that $n(n-1)/2$ is precisely the number of independent components of a $n$-dimensional 2-form, in the context of the general theory of relativity, B. Coll \cite{tolo} has conjectured that any $n$-dimensional metric $g$ can be locally obtained as a {\em deformation} of a constant curvature metric $\eta$, parametrized by a 2-form $F$, according to: 
\begin{equation} \label{e1}
 g = \lambda(F)\, \eta + \mu(F) F^2
\end{equation} 
where $\lambda$ and $\mu$ are scalar functions of $F$ and $F^2:=F\eta^{-1}F$.

It is worth noticing that the Kerr-Schild class of metrics in general relativity meet this relation. Indeed, a Kerr-Schild metric can be written as: $ g_{\alpha\beta} = \eta_{\alpha\beta} + l_\alpha l_\beta $, where $\eta_{\alpha\beta}$ is the Minkowski metric and $l_\alpha$ is a $\eta$-null vector field. Let $p_\alpha$ be a unit space vector field that is orthogonal to $l_\alpha$ and take $F_{\alpha\beta}= l_\alpha p_\beta - p_\alpha l_\beta$. Then the relation (\ref{e1}) holds with $\lambda=-\mu=1$.

In the present paper we prove the following variant of Coll's conjecture for $n=4$:

\medskip
\noindent {\bf Theorem 2 [deformation theorem]} \hspace*{1em} {\em Let $({\cal V}_n,g)$ be a semi-Riemannian manifold. Locally it always exists a 2-form $F$ and a scalar function $a$ such that:
\begin{list}
{(\alph{llista})}{\usecounter{llista}}
\item {they meet a previously chosen arbitrary scalar constraint:  $\;\Psi(a,F)=0$ and}
\item {the semi-Riemannian deformed metric:
 \begin{equation}\label{1a}
\overline{g}_{\alpha\beta} = a g_{\alpha\beta} -\epsilon F^2_{\alpha\beta} \;, 
\end{equation}
with  $F^2_{\alpha\beta}:= g^{\mu\nu}F_{\alpha\mu}F_{\nu\beta}$ and $|\epsilon| = 1$, has constant curvature.}
\end{list}  }

\noindent
Since the constraint $\Psi(a,F)$ is a scalar, it will only depend on the invariants of $F^\alpha_{\;\beta}=g^{\alpha\nu}F_{\nu\beta}$. 

As our proof relies on the application of the Cauchy-Kowalevski existence theorem for partial differential systems, it only applies to the analytic case. However, an extension to the ${\cal C}^\infty$ might be devised following the lines proposed by DeTurk and Yang \cite{Diag2} for a similar, although simpler, problem. 

\medskip \addtocounter{theorem}{1}
Theorem 2 is proved by iteration on the number of dimensions. Right from the start, it is true for $n=1$, because in one dimension any metric is flat, and for $n=2$ it reduces to the above mentioned Gauss theorem.

For the sake of iteration's convenience, what we actually proof in section 3 is the following extension of theorem 2:
\begin{theorem} \label{t2}
Let $({\cal V}_d,g)$ be a semi-Riemannian manifold and  $\lambda$ a 2-covariant symmetric tensor. Locally it always exists a 2-form $F$ and a scalar function $a$ such that:
\begin{list}
{(\alph{llista})}{\usecounter{llista}}
\item they meet a previously chosen constraint:  $\Psi(a,F,x)=0$ and
\item the semi-Riemannian deformed metric: 
\begin{equation}\label{1} \overline{g}_{\alpha\beta} = a
g_{\alpha\beta} +\lambda_{\alpha \beta} -\epsilon F^2_{\alpha\beta} \;,
\qquad{\rm with} \qquad  F^2_{\alpha\beta}:= g^{\mu\nu}F_{\alpha\mu}F_{\nu\beta} 
\end{equation} 
has constant curvature ($|\epsilon| = 1$).
\end{list}
\end{theorem}

A very useful tool for the iteration is the following

\smallskip\noindent
{\bf Theorem 4 [iteration]} \hspace*{1em} {\em If theorem \ref{t2} holds for $d=n-1$, then it also holds for $d=n$.}

\medskip\noindent\addtocounter{theorem}{1}
Section 2 is devoted to prove Theorem 4. Theorem 3 then follows by iteration and, as an application of theorem 2, Coll's conjecture is proved for $n=4$.

\section{Proof of the iteration theorem \label{S2}}
Let $({\cal V}_n,g)$ be a semi-Riemannian manifold and let $\{e_1, \ldots,e_n\}$ be a base of vectors. (Most tensor expressions in this proof are meant referred to this base.)

Both metrics $g_{\alpha\beta}$ i $\overline{g}_{\alpha\beta}$ will coexist through the proof. To avoid confusion, indices are always lowered with the metric $g_{\alpha\beta}$ and raised with its inverse $g^{\alpha\beta}$. Furthermore, $\overline h^{\alpha\beta}$ will denote the inverse metric  for $\overline{g}_{\alpha\beta}$ (see Appendix A).

Given the Riemannian connections $\nabla$ i $\overline{\nabla}$ respectively associated with $g$ i $\overline{g}$, we define the difference tensor 
\begin{equation}\label{B}
B^\alpha_{\mu\nu} := \overline{\gamma}^\alpha_{\mu\nu} - \gamma^\alpha_{\mu\nu}\,,
\end{equation}
so that 
\begin{equation} \label{3}  
\overline{\nabla}_\alpha A_\beta = \nabla_\alpha A_\beta - B^\mu_{\alpha\beta} A_\mu
\end{equation}
And it straightforwardly follows from the definition (\ref{B}) that
\begin{eqnarray}\label{4}
\overline B_{\rho\|\mu\nu} :=B^\alpha_{\mu\nu}\overline{g}_{\alpha\rho}&=& \frac12 \left(
\nabla_\mu\overline{g}_{\rho\nu}+\nabla_\nu\overline{g}_{\rho\mu}- \nabla_\rho\overline{g}_{\mu\nu}\right.)
\end{eqnarray}

Associated with each metric $g_{\alpha\beta}$ and $\overline{g}_{\alpha\beta}$ there is a Riemann tensor, respectively, $R_{\mu\nu\alpha\beta}$ and $\overline{R}_{\mu\nu\alpha\beta}$. Both tensors are related to each other by
\begin{equation}  \label{5}
\overline{R}_{\mu\nu\alpha\beta}:=\overline{g}_{\mu\rho} \overline{R}^\rho_{\;\nu\alpha\beta } =
\overline{g}_{\mu\rho}R^\rho_{\;\nu\alpha\beta } + 2\,\nabla_{[\alpha}\overline B_{\mu\|\beta ]\nu} + 2\,B^\rho_{\nu[\alpha}\overline B_{\rho\|\beta ]\mu}
\end{equation}

This equation, together with the condition that $\overline{g}_{\alpha\beta}$ has constant curvature, yields the partial differential system 
\begin{equation} \label{6} 
{\cal E}_{\mu\nu\alpha\beta }:= \overline{R}_{\mu\nu\alpha\beta }+k
\overline{g}_{\mu\nu\alpha\beta }=0 
\end{equation} 
where $\overline{g}_{\mu\nu\alpha\beta}:= \overline{g}_{\mu\alpha}\overline{g}_{\nu\beta}-\overline{g}_{\mu\beta}\overline{g}_{\nu\alpha}$. It is a system with $1+n(n-1)/2$ unknowns, namely, the scalar function $a$ and the independent components of $F_{\alpha\beta}$. On the other hand, considering the symmetries of the Riemann tensor, the system (\ref{6}) consists of $n^2(n^2-1)/12$ independent equations \cite{Weinberg}. Therefore, for $n>2$ there are more equations
than unknowns, and the system seems overdetermined. (We shall see that this overdetermination is similar to what one encounters in analyzing Einstein equations: part of equations (\ref{6}) constitute a partial differential system with a well posed Cauchy problem, while the remaining equations are subsidiary conditions on the Cauchy data.)

\subsection{The Cauchy problem \label{sec:2.1}}
A suitably chosen part of equations (\ref{6}) yields a partial differential system for the unknowns $F_{\alpha\beta}$ and $a$. To pose the corresponding Cauchy problem, we shall take a hypersurface ${\cal S}$ with a non-null normal vector $n^\alpha$, that we assume normalized, i.e. $g_{\alpha\beta} n^\alpha n^\beta =\sigma$\,, $|\sigma|= 1$\,. We shall consider an extension of $n^\alpha$ to a neighborhood of ${\cal S}$, which will
be denoted by the same symbol, and 
\begin{equation} \label{7}
 \Pi^\alpha_\mu := \delta^\alpha_\mu -\sigma n^\alpha n_\mu
\end{equation} 
is the projector $g$-orthogonal to $n^\alpha$.

It is easily seen that not all equations  (\ref{6}) contain second order normal derivatives of the unknowns. In particular, the only equations contributing to the principal part \cite{John} of the system are the combinations: 
\begin{equation} \label{8} 
{\cal E}_{\mu\alpha}:= {\cal E}_{\mu\nu\alpha\beta}\,n^\nu n^\beta =0 
\end{equation} 
and, considering the identities 
\begin{equation}\label{9} 
{\cal E}_{\mu\alpha} \equiv {\cal E}_{\alpha\mu} \; , \qquad {\cal E}_{\mu\alpha}n^\mu \equiv 0 
\end{equation} 
it is obvious that (\ref{8}) consists of  $n(n-1)/2$ independent equations (as many
as unknowns but one). As it will be shown later on, a well posed Cauchy problem with data on the hypersurface ${\cal S}$ can be set up for these equations.

As for the remaining equations (\ref{6}), they can be gathered as
\begin{eqnarray}
{\cal L}_{\mu\nu\alpha\beta } :={\cal E}_{\lambda\rho\sigma\gamma}\,
\Pi^\lambda_\mu\, \Pi^\rho_\nu\, \Pi^\sigma_\alpha\, \Pi^\gamma_\beta  =0 \label{10}\\
{\cal L}_{\mu\nu\alpha} := {\cal E}_{\lambda\rho\sigma\gamma}\,\Pi^\lambda_\mu\, \Pi^\rho_\nu\,
\Pi^\sigma_\alpha\, n^\gamma  =0\label{11} 
\end{eqnarray} 
where $\Pi^\lambda_\mu$ is the projector defined in (\ref{7}). These equations can be taken as subsidiary conditions to be fulfilled by the Cauchy data on ${\cal S}$. Indeed:

\begin{proposition}\label{p1} Any analytical solution of (\ref{8}) that meets the constraints (\ref{10}) and (\ref{11}) on ${\cal S}$, also meets these conditions in a neighborhood of ${\cal S}$. 
\end{proposition}

\paragraph{Proof:} Using the combinations (\ref{8}), (\ref{10}) and (\ref{11}), we can write: 
\begin{equation} \label{12} 
{\cal E}_{\mu\nu\alpha\beta }\equiv {\cal L}_{\mu\nu\alpha\beta }  + 2
\sigma \left({\cal L}_{\mu\nu[\alpha}n_{\beta] }+{\cal
L}_{\alpha\beta[\mu}n_{\nu] }\right) + 4 n_{[\nu}{\cal E}_{\mu][\alpha}
n_{\beta]} 
\end{equation} 
Hence, given a solution of (\ref{8}) we have that:
\begin{equation} \label{12a} 
{\cal E}_{\mu\nu\alpha\beta }={\cal L}_{\mu\nu\alpha\beta }  + 2 \sigma \left({\cal
L}_{\mu\nu[\alpha}n_{\beta] }+{\cal L}_{\alpha\beta[\mu}n_{\nu]
}\right) 
\end{equation}

Since the Riemann tensor $\overline{R}_{\mu\nu\alpha\beta }$ meets the second Bianchi identity for the connection $\overline\nabla$ and $\overline\nabla_\alpha \overline g_{\mu\nu}=0$, the following identity holds:
\begin{equation}
 \overline{\nabla}_\lambda {\cal E}_{\mu\nu\alpha\beta } +
 \overline{\nabla}_\alpha {\cal E}_{\mu\nu\beta\lambda } +
 \overline{\nabla}_\beta {\cal E}_{\mu\nu\lambda\alpha } \equiv 0
\label{12b}
\end{equation}
which, with the help of the difference tensor (\ref{B}), can be written as:
\begin{equation}\label{13} 
\nabla_\lambda {\cal E}_{\mu\nu\alpha\beta} + \nabla_\alpha {\cal E}_{\mu\nu\beta\lambda} + \nabla_\beta {\cal E}_{\mu\nu\lambda\alpha} + {\rm linear\,} ({\cal E}) \equiv 0 
\end{equation} 
where ``linear$({\cal E})$'' stands for terms that depend linearly on ${\cal E}_{\mu\nu\alpha\beta }$. In particular, by substituting (\ref{12a}) into (\ref{13}) and contracting with $n^\lambda \Pi^\alpha_\rho \Pi^\beta_\sigma$, we obtain
\begin{eqnarray*}\label{14}
&\nabla_n {\cal L}_{\mu\nu\rho\sigma} + 2\sigma \nabla_n
{\cal L}_{\rho\sigma[\mu}n_{\nu]} + \Pi^\alpha_{\rho}  \nabla_\alpha
{\cal L}_{\mu\nu\sigma} -\Pi^\beta_{\sigma}  \nabla_\beta
{\cal L}_{\mu\nu\rho} + {\rm linear\,} ({\cal L}) = 0
\end{eqnarray*}
which, projected on $n^\nu$, yields: 
\begin{equation}\label{15} 
\nabla_n {\cal L}_{\rho\sigma\mu}+ {\rm linear\,} ({\cal L}) = 0
\end{equation}
and, contracted with  $\Pi^\mu_\tau\Pi^\nu_\lambda$ yields: 
\begin{equation}\label{16} 
\nabla_n {\cal L}_{\tau\lambda\rho\sigma}+ \Pi^\alpha_{\rho}\nabla_\alpha {\cal L}_{\tau\lambda\sigma} -\Pi^\beta_{\sigma} \nabla_\beta {\cal L}_{\tau\lambda\rho} + {\rm linear\,} ({\cal L}) = 0 
\end{equation}

Equations (\ref{15}) i (\ref{16}) constitute a homogeneous, linear, partial differential system for the unknowns ${\cal L}_{\tau\lambda\rho\sigma}$ and ${\cal L}_{\rho\sigma\mu}$. The hypersurface ${\cal S}$ is clearly non-characteristic for this system and, since ${\cal L}_{\tau\lambda\rho\sigma}$ and ${\cal L}_{\rho\sigma\mu}$ vanish on ${\cal S}$, they will also vanish on a neighborhood.\footnote{The validity of this reasoning, which relies on a uniqueness theorem, is thus restricted to the analytic case (see \cite{John}).} \hfill $\Box$

\subsection{Non-characteristic hypersurfaces}
Let us now see that Cauchy data can be provided so that ${\cal S}$ is non-characteristic for the partial differential system (\ref{8}). We first use (\ref{4}), (\ref{5}) and (\ref{9}) to write the system as: 
\begin{equation}\label{2.1} 
{\cal E}_{\alpha\beta} = -\frac12 \nabla_n^2\left(\Pi^\mu_\alpha\Pi^\nu_\beta \overline{g}_{\mu\nu}\right) + {\rm n.\,p.\,t\,} =0 
\end{equation} 
where ``n.\,p.\,t\,'' includes all non-principal terms, i.e. terms that do not contain second order normal derivatives of the unknowns.

We then consider the splitting of $F_{\alpha\beta}$ into its parallel and transversal parts with respect to $n_\alpha$\,:
\begin{equation}\label{2.2} 
F_{\alpha\beta}:= n_\alpha E_\beta - E_\alpha n_\beta + \tilde{F}_{\alpha\beta} 
\end{equation} 
where
$$E_\alpha:=-\sigma F_{\alpha\mu}n^\mu  \qquad{\rm and} \qquad \tilde{F}_{\alpha\beta}:= \Pi^\mu_\alpha\Pi^\nu_\beta F_{\mu\nu}\;.$$
Using this, the combination $\Pi^\mu_\alpha\Pi^\nu_\beta\overline{g}_{\mu\nu}$ occurring in the principal terms in equation (\ref{2.1}) can be written as
\begin{equation} \label{2.2a}
\Pi^\mu_\alpha\Pi^\nu_\beta\overline{g}_{\mu\nu} = a \, \hat g_{\alpha\beta}  + \Pi^\mu_\alpha\Pi^\nu_\beta \lambda_{\mu\nu} +\sigma\epsilon \,E_\alpha E_\beta - \epsilon\, \tilde{F}_{\alpha}^{\;\rho} \,\tilde{F}_{\rho\beta}
\end{equation}
which, substituted into (\ref{2.1}), yields:
\begin{equation} \label{2.2b}
\ddot a \, \hat g_{\alpha\beta}  + \sigma\epsilon \,\left(\ddot
E_\alpha E_\beta + E_\alpha \ddot E_\beta \right) - \epsilon\,
\left(\ddot{\tilde{F}}_{\alpha}^{\;\rho} \,\tilde{F}_{\rho\beta} +
\tilde{F}_{\alpha}^{\;\rho} \,\ddot{\tilde{F}}_{\rho\beta} \right)
= {\cal H}_{\alpha\beta}
\end{equation}
where ${\cal H}_{\alpha\beta}$ includes all non-principal terms and ``$(\,\ddot{ }\,)$'' stands for $\nabla_n^2$. Notice that, since $\lambda_{\alpha\beta}$ is given, its derivatives do not contribute the principal part.

Recall now that there is a beforehand fixed arbitrary relation among the coefficient $a$ and the components $F_{\alpha\beta}$, 
\begin{equation} \label{2.2c} 
\Psi(a,F_{\rho\beta},x) = 0 
\end{equation}
which implies the following constraint on the second normal derivatives of the unknowns: \begin{equation} \label{2.2d} 
\Psi_1\ddot a + \Psi_2^\alpha \ddot E_\alpha +
\Psi_3^{\alpha\beta}\ddot{\tilde{F}}_{\alpha\beta} = \Psi_0 \end{equation}
where
$$\Psi_1:=\frac{\partial\Psi}{\partial a}\;, \qquad
\Psi_2^\alpha:=\frac{\partial\Psi}{\partial E_\alpha} \qquad {\rm
and} \qquad \Psi_3^{\alpha\beta}:=\frac{\partial\Psi}{\partial
\tilde{F}_{\alpha\beta}}  $$ 
are the partial derivatives of $\Psi$ relatively to each of its independent variables, and
$\Psi_0$ includes all non-principal terms.

The hypersuface ${\cal S}$ is non-characteristic for the prescribed Cauchy data if, and only if, the linear system (\ref{2.2b}-\ref{2.2d}) can be solved for the second order
normal derivatives $\ddot E_\alpha$, $\ddot{\tilde{F}}_{\alpha\beta}$ and $\ddot a$. In Appendix C a sufficient condition on the Cauchy data $E_\alpha$ and $\tilde{F}_{\alpha\beta}$ is derived for ${\cal S}$ to be non-characteristic, namely 
\begin{equation} \label{2.4}
 \Delta(d) \neq 0 \qquad {\rm and} \qquad \Gamma(d) \neq 0
\end{equation} 
where $\Delta(d)$ is the determinant of the system (\ref{A23a}-\ref{A23b}) and $\Gamma(d)$ is the determinant of $G_{ij}:= (-1)^{i-1} E_\alpha E_\beta (\tilde{F}^{i+j-2})^{\alpha\beta} $, $i,j = 1 \ldots d$. (These conditions are equivalent to the inequalities (\ref{A31}).) Also in Appendix C, explicit expressions of these conditions are listed for some few values of $d$, the number of dimensions of the Cauchy hypersurface ($d=1,2,3$).

That is, it suffices to pick a set of Cauchy data on ${\cal S}$ fulfilling the inequalities  (\ref{2.4}). Notice that these only involve the values of $a$, $E_\alpha$ and $\tilde{F}_{\alpha\beta}$ on ${\cal S}$, but not their normal derivatives.

\subsection{Geometrical meaning of the subsidiary conditions \label{sec2.2}}
Let us now assume that the hypersurface ${\cal S}$ has been chosen so that $\overline{h}^{\alpha\beta}n_\alpha n_\beta\neq 0$. We can then 
consider the unit vector $\overline{g}$-orthogonal to ${\cal S}$, which is obtained on raising the index in $n_\beta$ with $\overline{h}^{\alpha\beta}$ and normalizing: 
\begin{equation} \label{3.1} 
\zeta_\beta:= \xi n_\beta \;, \qquad \xi:=|\overline{h}^{\alpha\beta}n_\alpha n_\beta|^{-1/2} \;, \qquad  \overline{n}^\alpha:= \overline{h}^{\alpha\beta}
\zeta_\beta 
\end{equation} 
and therefore,  $\overline{g}_{\alpha\beta} \overline{n}^\alpha \overline{n}^\beta = \bar\sigma\,$, $|\bar\sigma|= 1$.

We denote 
\begin{equation} \label{3.2} 
\overline\Pi^\alpha_\beta := \delta^\alpha_\beta - \bar\sigma \overline{n}^\alpha \zeta_\beta
\end{equation} 
the projector $\overline{g}$-orthogonal to $\overline{n}^\alpha$. Since, by definition $n_\alpha$ and $\zeta_\alpha$ are proportional to each other, we have that
$$ n_\alpha \overline\Pi^\alpha_\beta = \zeta_\alpha \Pi^\alpha_\beta = 0 $$
Whence it follows straightforwardly that 
\begin{equation} \label{3.3}
\Pi^\alpha_\beta \overline\Pi^\beta_\mu = \overline\Pi^\alpha_\mu
\qquad {\rm i} \qquad \overline\Pi^\alpha_\beta \Pi^\beta_\mu =
\Pi^\alpha_\mu 
\end{equation}

\begin{proposition}
The subsidiary conditions (\ref{10}) and (\ref{11}) are equivalent to: 
\begin{equation} \label{3.4} 
\left. \begin{array}{l}
\overline{\cal L}_{\mu\nu\alpha\beta } :={\cal E}_{\lambda\rho\sigma\gamma}\, \overline\Pi^\lambda_\mu\, \overline\Pi^\rho_\nu\, \overline\Pi^\sigma_\alpha\, \overline\Pi^\gamma_\beta  =0 \qquad \\
\overline{\cal L}_{\mu\nu\alpha} :=
{\cal E}_{\lambda\rho\sigma\gamma}\,\overline\Pi^\lambda_\mu\,
\overline\Pi^\rho_\nu\, \overline\Pi^\sigma_\alpha\, \overline{n}^\gamma  =0
\end{array}  \right\}
\end{equation}
\end{proposition}

\paragraph{Proof:} Indeed, conditions (\ref{10}) and (\ref{11}) are met if, and only if,  ${\cal E}_{\lambda\rho\sigma\beta} \,\Pi^\lambda_\mu\, \Pi^\rho_\nu\, \Pi^\sigma_\alpha  =0 \,$, which implies that 
$$ {\cal E}_{\lambda\rho\sigma\beta}\,\Pi^\lambda_\mu\, \Pi^\rho_\nu\, \Pi^\sigma_\alpha \,\overline\Pi^\mu_\kappa\, \overline\Pi^\nu_\tau\, \overline\Pi^\alpha_\gamma = 0 $$
whence, including (\ref{3.3}), it follows that: ${\cal E}_{\lambda\rho\sigma\beta} \,\overline\Pi^\lambda_\kappa\, \overline\Pi^\rho_\tau\, \overline\Pi^\sigma_\gamma  =0$, which  leads straightforwardly to (\ref{3.4}). The converse proof proceeds in a similar way. \hfill $\Box$

Written in the form (\ref{3.4}) the subsidiary conditions have a clear geometrical meaning. We only need to notice that $\overline{g}_{\alpha\beta}$ is a semi-Riemannian metric and ${\cal S}$ is a hypersurface. The theory of Riemannian submanifolds can be then invoked \cite{Hicks},\cite{Choquet}. Indeed, given ${\cal S}$ we can consider either the isometric embedding in the Riemannian manifold $({\cal V}_4,g)$ or the isometric embedding in the Riemannian manifold $({\cal V}_4,\overline g)$. For each embedding we have a first fundamental form, respectively: 
\begin{equation} \label{3.9} 
\hat g_{\alpha\beta} := g_{\mu\nu} \Pi^\mu_\alpha\,\Pi^\nu_\beta \qquad{\rm and} \qquad
\tilde g_{\alpha\beta} := \overline g_{\mu\nu} \overline \Pi^\mu_\alpha\,\overline \Pi^\nu_\beta 
\end{equation}
where $\Pi^\mu_\alpha$ and $\overline \Pi^\mu_\alpha$ are the respective projectors. We also have a second fundamental form for each embedding, respectively: 
\begin{equation} \label{3.10} 
\phi_{\alpha\beta}:= \Pi^\mu_\alpha\,\Pi^\nu_\beta\,\nabla_\mu n_\nu \qquad{\rm and}
\qquad \overline \phi_{\alpha\beta}:= \overline \Pi^\mu_\alpha\,\overline \Pi^\nu_\beta\, \overline\nabla_\mu \overline n_\nu 
\end{equation}

Then, with the help of the projector $\overline\Pi^\lambda_\alpha$ and the connection $\overline\nabla$, a connection $\tilde\nabla$ can be defined on ${\cal S}$, which is precisely the Riemannian connection for $\tilde{g}_{\alpha\beta}$. Let $v^\alpha$ and
$w^\beta$ be two  vector fields tangential to ${\cal S}$, then:
$$\tilde\nabla_v w^\alpha := \overline\Pi^\alpha_\beta\,\overline\nabla_v w^\beta $$
and also:
$$\overline\phi_{\alpha\beta} v^\alpha w^\beta = \overline\nabla_v \zeta_\beta\, w^\beta = -\overline\nabla_v w^\beta \,\zeta_\beta $$

The subsidiary conditions (\ref{3.4}) can be geometrically interpreted in the light of the Gauss and Codazzi-Mainardi equations \cite{Hicks} for the submanifold ${\cal S}$.

\subsubsection{The Codazzi-Mainardi equation \label{sec2.3}}
It is basically a relation between some components of the Riemann tensor  $\overline{R}_{\mu\nu\alpha\beta}$ and the derivatives of the second fundamental form \cite{Hicks}. Let $v^\mu$, $w^\nu$ and $z^\alpha$ be three vectors tangential to
${\cal S}$, then:
$$ \overline{R}_{\mu\nu\alpha\beta}v^\mu w^\nu z^\alpha \overline{n}^\beta = \overline{R}(v,w,z,\overline{n}) =\tilde\nabla_v \overline\phi(w,z) - \tilde\nabla_w \overline\phi(v,z) $$
Therefore, the second subsidiary condition (\ref{3.4}) is equivalent to: 
\begin{equation} \label{3.7} 
\tilde\nabla_v \overline\phi(w,z) - \tilde\nabla_w \overline\phi(v,z) = 0 
\end{equation} 
for any $v$, $w$ and $z$ tangential to ${\cal S}$.

A particular solution is: 
\begin{equation} \label{3.7a} 
\overline\phi(v,w) = 0 \qquad \forall \,v,\, w \quad\mbox{tangential to }\,{\cal S} 
\end{equation}
which can always be achieved by a suitable choice of the normal derivatives of the unknowns on ${\cal S}$. Indeed, substituting (\ref{3.1}) in the first expression (\ref{3.10}) and including (\ref{3}), we obtain:
$$ \xi \phi_{\alpha\beta} = \Pi^\lambda_\alpha\, \Pi^\mu_\beta \,
 \left[\overline\nabla_\lambda \overline n_\mu +B^\tau_{\lambda\mu} \overline n_\tau\right] $$
which, using the relations (\ref{3.3}) and (\ref{4}), leads to 
\begin{eqnarray*}
\xi \phi_{\alpha\beta} &=& \Pi^\lambda_\alpha\, \Pi^\mu_\beta \overline\phi_{\lambda\mu} + \overline n^\tau\nabla_\lambda \overline g_{\tau\mu} \Pi^\lambda_{(\alpha} \Pi^\mu_{\beta)} \\
   & & - \frac12 \overline n^\tau \Pi^\rho_\tau \nabla_\rho \overline g_{\lambda\mu} \Pi^\lambda_{\alpha} \Pi^\mu_{\beta} - \frac\sigma2 (n\overline n)\,\Pi^\lambda_{\alpha} \Pi^\mu_{\beta}\,\nabla_n \overline g_{\lambda\mu}
\end{eqnarray*}
Finally, according to (\ref{3.1}), we have that $n_\alpha \overline{n}^\alpha = \xi^{-1}\bar\sigma$ and the condition $\overline\phi_{\alpha\beta}=0$ can be written as \begin{equation}
\label{3.11} \nabla_n \left[\overline
g_{\lambda\mu}\,\Pi^\lambda_{\alpha}\, \Pi^\mu_{\beta}\right] ={\cal K}_{\alpha\beta} 
\end{equation} 
where the right hand-side only depends on the values of $\overline g_{\alpha\beta}$ on  ${\cal S}$ and its tangential derivatives.

The relations (\ref{3.11}) are $d(d+1)/2$ independent equations. Complemented with the constraint obtained on differentiating (\ref{2.2c}) once 
\begin{equation} \label{3.12} 
\Psi_1\dot a + \Psi_2^\alpha \dot E_\alpha + \Psi_3^{\alpha\beta} \dot{\tilde{F}}_{\alpha\beta} = \Psi_4 
\end{equation} 
we obtain a linear system with $1+d(d+1)/2$ equations that can be solved for the normal derivatives $\dot a$, $\dot E_\alpha$ and $\dot{\tilde{F}}_{\alpha\beta}$, provided that
the  determinant does not vanish ($\Psi_4$ only depends on $a$, $E_\alpha$, $\tilde{F}_{\alpha\beta}$ and their tangential derivatives). Note also that the structure of the linear system (\ref{3.11})-(\ref{3.12}) is the same as the system (\ref{2.2b})-(\ref{2.2d}), therefore it has a unique solution for $\dot a$, $\dot E_\alpha$ i $\dot{\tilde{F}}_{\alpha\beta}$ whenever the inequalities (\ref{2.4}) are met.

\subsubsection{The Gauss equation \label{sec2.4}}
The Gauss equation relates the Riemann tensor for the metric $\tilde{g}$ and the tangential components (i. e., on projecting with $\overline\Pi^\alpha_\beta$) of the Riemann tensor for the metric $\overline{g}$ \cite{Hicks}. So, if the vectors $v$, $w$, $t$ and
$z$ are tangential to ${\cal S}$, then 
$$\overline{R}(v,w,t,z) = \tilde{R}(v,w,t,z) -\bar\sigma\left[\overline \phi(v,t)
\overline\phi(w,z) -\overline\phi(v,z)\overline\phi(w,t)\right] \,. $$
Now, the first subsidiary condition (\ref{3.4}) is:
$$ {\cal E}_{\mu\nu\alpha\beta }v^\mu w^\nu t^\alpha z^\beta:=\overline{R}(v,w,t,z)+
k\, \left[\overline{g}(v,t) \overline{g}(w,z) -\overline{g}(v,z)\overline{g}(w,t)\right]=0 \,. $$
On combining both equations and introducing the particular choice (\ref{3.7a}), we obtain:
\begin{equation} \label{3.8}
\tilde{R}(v,w,t,z)+ k\,
\left[\overline{g}(v,t) \overline{g}(w,z) -\overline{g}(v,z)\overline{g}(w,t)\right]=0
\end{equation}

Let us now consider a base of vectors adapted to ${\cal S}$, i.e. $\{t_1^\alpha\,\ldots t_d^\alpha, t_n^\alpha=n^\alpha\}$, with $t_j^\alpha n_\alpha=t_j^\alpha \zeta_\alpha =0$\,, $j=1 \ldots d$. It is obvious that $\Pi^\alpha_\mu t_j^\mu = t_j^\alpha$.

The metric $\tilde{g}$ on ${\cal S}$ has the components:
$$ \tilde{g}_{ij} = \tilde{g}_{\alpha\beta} t_i^\alpha t_j^\beta =
\left(\overline{g}_{\mu\nu} \,\overline\Pi^\mu_\alpha\,\overline\Pi^\nu_\beta \right)\,
\Pi^\alpha_\rho t_i^\rho\, \Pi^\beta_\lambda t_j^\lambda $$
and using the second relation (\ref{3.3}) we arrive at:  $\tilde{g}_{ij} = \overline{g}_{\alpha\beta}
\, \Pi^\alpha_\rho \, \Pi^\beta_\lambda t_i^\rho  t_j^\lambda \;$ which, including
(\ref{2.2a}), allows to write 
\begin{equation} \label{2.41}
\tilde{g}_{ij} = a\,\hat g_{ij}+ \tilde\lambda_{ij} + \sigma\epsilon E_i E_j  - \epsilon
\tilde{F}^2_{ij} \;, \qquad i,j=1 \ldots d
\end{equation}
with $\tilde\lambda_{ij} := \lambda_{ij} + \sigma\epsilon E_i E_j $. Also in this base, equation (\ref{3.8}) reads: 
$$ \tilde{R}_{ijkl} + k \left[\tilde{g}_{ik}\tilde{g}_{jl}- \tilde{g}_{il}\tilde{g}_{jk}\right] =0 \;, \qquad i,j,k,l=1 \ldots d $$
i. e., the Cauchy data on  ${\cal S}$ must be chosen so that the metric $ \tilde{g}_{ij}$ has constant curvature.

Now, seeking a 2-form $\tilde{F}_{ij}$ and a coefficient $a$ such that the metric $\tilde{g}_{ij}$ has constant curvature, is a problem similar to what we were facing through this section (but with a lesser number of dimensions $d=n-1$) which, by the hypothesis of theorem 4, has already a solution\footnote{The constraint $\tilde\Psi$ for this new reduced problem is $\tilde\Psi(a,\tilde{F},x)\equiv\Psi(a,n_\alpha E_\beta-E_\alpha n_\beta+\tilde{F}_{\alpha\beta},x)$. Notice that, although we had taken $\Psi$ independent of the point $x$, as in Theorem 2, $\tilde\Psi$ would indeed depend on $x$ through $E_\alpha$ and $n_\beta$.}. 

Summarizing, we choose a hypersurface ${\cal S}\subset {\cal V}_d$ and find a set of Cauchy data on ${\cal S}$,  $ \{a ,\; E_\alpha , \; \tilde{F}_{\alpha\beta} \}\,$ such that:
\begin{list}
{(\alph{llista})}{\usecounter{llista}}
\item $ E_\alpha n^\alpha =0 \,$, and $\tilde{F}_{\alpha\beta} n^\beta=0 \,$,
$\alpha,\beta=1,\ldots,d$,
\item $\Delta(d)\neq 0$ and  $\Gamma(d)\neq 0$  (see Appendix C) and
\item the metric $\tilde{g}_{ij} = a\,\hat g_{ij}+ \tilde\lambda_{ij} - \epsilon \tilde{F}^2_{ij} \,$ on ${\cal S}$ has constant curvature.
\end{list}
Then the Cauchy data are completed on solving the system (\ref{3.11}-\ref{3.12})
for $\dot a$, $\dot E_\alpha$ and $\dot{\tilde{F}}_{\alpha\beta}$.

With the complete set of Cauchy data, we solve the partial differential system (\ref{8})
to obtain $a ,\; E_\alpha , \; \tilde{F}_{\alpha\beta}$ in a neighborhood of ${\cal S}$.  Then, using (\ref{2.2}) and (\ref{1a}), the sought constant curvature metric $\overline{g}_{\alpha\beta}$ is finally obtained.  \hfill $\Box$

\section{Proof of Theorem 3 \label{S3}}
Theorem 3 is then proved by iteration on the number of dimensions. First, the theorem holds for $n=1$, because in that case every metric is flat. And second, by Theorem 4, if the statement of Theorem 3 holds for $n-1$, then it holds for $n$ too. The proof then follows as an application of the recurrence principle.

Finally, Theorem 2 is a particular case of Theorem 3 for $\lambda_{\alpha\beta} =0$ and  $\Psi$ independent of $x$.

\section{Application. A universal law of gravitational deformation\label{S4}}
We shall now apply Theorem 2 to prove Coll's conjecture \cite{tolo} for a  Lorentzian, 4-dimensional spacetime, which we state as
\begin{theorem}
Every 4-dimensional Lorentzian metric $g_{\alpha\beta}$ can be locally written as:
\begin{equation} \label{4.1}
 g_{\alpha\beta}=\lambda\,\overline{g}_{\alpha\beta} + \mu\, \overline{F}^{2}_{\alpha\beta}
\end{equation}
where:
\begin{list}
{(\alph{llista})}{\usecounter{llista}}
\item $F_{\alpha\beta}$ is a 2-form and $\lambda$ and $\mu$ are scalar functions of $F_{\alpha\beta}$,
\item  $\overline{g}_{\alpha\beta}$ is a constant curvature Lorentzian metric and
\item  \begin{equation}  \label{4.2}
 \overline{F}^{2}_{\alpha\beta}:=F_{\alpha\mu}\overline{h}^{\mu\nu} F_{\nu\beta} \,,
\end{equation} 
where $\overline{h}^{\mu\nu}$ is the inverse metric for $\overline{g}_{\alpha\beta}$.
\end{list}
\end{theorem}

\medskip\noindent
{\bf Proof:\hspace*{1em}} By Theorem 2, a 2-form $F_{\alpha\beta}$ and a scalar function $a$ exist such that 
\begin{equation} \label{4.3} 
\overline{g}_{\alpha\beta} = a g_{\alpha\beta} -\epsilon F^2_{\alpha\beta} \;, \qquad{\rm
with} \qquad  F^2_{\alpha\beta}:= F_{\alpha\mu} g^{\mu\nu}F_{\nu\beta} 
\end{equation}
has constant curvature.

Now, solving (\ref{4.1}) for $\overline{g}_{\alpha\beta}$, we have $\lambda\, \overline{g}_{\alpha\beta} =  g_{\alpha\beta} -{\mu} F_{\alpha\mu} \overline{h}^{\mu\nu} F_{\nu\beta} \;$ which, using (\ref{A11}) and (\ref{A12}) (Appendix B) becomes:
$$\overline{g}_{\alpha\beta} = \frac1\lambda\,\left(1+\frac{\epsilon\mu I_2}{p(\epsilon a)} \right)\,
g_{\alpha\beta} -\frac{a \mu }{\lambda p(\epsilon a)}\, F^2_{\alpha\beta} $$
where $I_1$ and $I_2$ are the coefficients of the characteristic polynomial $\chi_F(X):=\det(F^\alpha_{\;\beta}-X\delta^\alpha_{\;\beta})$ and $p(x):=x^2+I_1x+I_2$.

On comparing it with (\ref{4.3}), we arrive at
\begin{equation} \label{4.4}
a = \frac1\lambda\,\left(1+\frac{\epsilon\mu I_2}{p(\epsilon a)} \right) \qquad {\rm and} \qquad \epsilon = \frac{a \mu }{\lambda p(\epsilon a)}
\end{equation}
If we then solve these equations for $\lambda$ and $\mu$ we finally obtain:
\begin{equation} \label{4.5}
\lambda=\frac{a}{a^2+I_2}\qquad {\rm and} \qquad \mu = \epsilon +\frac{a I_1 }{a^2+I_2}  \,.
\end{equation}
Including then the scalar constraint $\Psi(a,F)=0$, which is an implicit equation connecting $a$, $I_1$ and $I_2$, it follows that $\lambda$ and $\mu$ only depend on the invariants $I_1$ and $I_2$. 

\section{Conclusion}
We have proved that, locally, any semi-Riemannian metric can be obtained as a deformation of a constant curvature metric according to the deformation law (\ref{1a}). As, once the constrain $\Psi(a,F)=0$ is fixed, the conformal factor $a$ can be written as a function of the 2-form $F$, a correspondence can be established between the space of semi-Riemannian metric $g_{\alpha\beta}$ and  the space of 2-forms $F_{\alpha\beta}$. We have also noticed that this kind of deformations includes Kerr-Schild transformations \cite{KerrSchild} and generalized Kerr-Schild transformations \cite{SenoHilde}.

In general relativity ($n=4$) this correspondence could open the possibility of alternative approaches to the classification of metrics and spacetimes, namely, on the basis of a classification of 2-forms.

It will also permit to solve the gravitational field equations for $F_{\alpha\beta}$, or for $M_{\alpha\beta}=F_{\alpha\mu} F^\mu_{\;\beta}$ as a perturbation of a flat metric $\overline{g}_{\alpha\beta}$ (or a constant curvature one, at the best convenience). This could provide a new useful tool to study exact solutions of Einstein equations.

Nevertheless, we must not forget the limitations of the result just proved, nor the great task that is still left from a more fundamental (mathematical) point of view. We have striven to make clear that our proof is only valid in the analytic case, i. e. if all data are real analytic functions. Therefore, an outstanding job is to extend the validity of our theorem to the case that the data are only ${\cal C}^\infty$ functions. An approach could consist of transforming the quasilinear partial differential system (\ref{2.2b}-\ref{2.2d}) so that it can be linearized into a symmetric hyperbolic system. In such a case the existence of a solution for ${\cal C}^\infty$ data would follow from the inverse function theorem of Nash and Moser \cite{NashMoser}. 

We have also insisted in the fact that our result is valid only locally, in a neighborhood of the Cauchy surface. It will be important to investigate how far the solution can be continued.

It must be also kept in mind that the deformation of a given metric by means of a 2-form, according to the rule (\ref{1a}), to obtain a constant curvature metric has not a unique solution. Besides the arbitrariness in the choice of the constraint $\Psi$, a great freedom is left in the choice of the Cauchy data on ${\cal S}$. To put it in a different way, think of the great variety of 2-forms $F$ that deform a given constant curvature metric $\eta$ into another metric $\overline{\eta}=\eta -\epsilon F^2$ which has the same constant curvature. The ``gauge'' problem concerning this class of deformations is still to be studied. 

\section*{Acknowledgment}
We are indebted to J. Carot, B. Coll and M. Mars for stimulating discussions and enlightening comments.

This research was supported by Ministerio de Ciencia y Tecnolog\'{\i}a, contract no. BFM2003-07076, and Generalitat de Catalunya, contract no. 2001SGR-00061 (DURSI).

\subsection*{Appendix A: The inverse metric $\overline{h}^{\alpha \beta}$}
We are seeking the inverse metric $\overline{h}^{\alpha \beta}$ for the covariant metric
\begin{equation} \label{A1} 
\overline{g}_{\alpha \beta} = a\,g_{\alpha\beta} - \epsilon\,M_{\alpha\beta} \qquad {\rm with} \qquad M_{\alpha\beta} :=F_{\alpha\mu} F^\mu_{\;\beta} 
\end{equation} 
and $\alpha,\beta,\mu = 1,\ldots n$.

Since $F_{\alpha\beta}$ is skewsymmetric, the characteristic polynomial $\chi_F(X):= \det(F^\mu_{\;\beta}-X \delta^\mu_{\;\beta})$ has degree $n$ and only involves powers $X^r$ with $n-r$ even. Denoting $\mathbb{F}=(F^\mu_{\;\beta})$, we can write:
$$ \chi_F(\mathbb{F}):=\mathbb{F}^n + I_1 \mathbb{F}^{n-2} + \ldots =0$$
Therefore, for the matrix $\mathbb{M}=\mathbb{F}^2$ we have that
$$\mathbb{M}^q + I_1 \mathbb{M}^{q-1} + \ldots =0 \qquad {\rm with} \qquad q=\left[\frac{n+1}2\right]$$
and the minimal polynomial for $\mathbb{M}$ has the form: 
\begin{equation}
\label{A2} 
m(\mathbb{M}):=\mathbb{M}^r + \sum_{j=0}^{r-1} k_j \mathbb{M}^j =0
\qquad {\rm with} \qquad r\leq\left[\frac{n+1}2\right] 
\end{equation}

Introducing now the matrix $\overline{\mathbb{G}}=(\overline{g}^\alpha_{\;\beta})$, equation (\ref{A1}) can be written as: $\mathbb{G}= a\,\mathbb{I}-\epsilon\,\mathbb{M}$ and, if we denote $\mathbb{H}=(\overline{h}^\alpha_{\;\beta})$, then $\overline{\mathbb{G}}\mathbb{H} =\mathbb{I}$, i. e. 
\begin{equation}
\label{A3}
 a\,\mathbb{H} - \epsilon\,\mathbb{M}\mathbb{H} = \mathbb{I}
\end{equation} 
We can find a solution in the shape: $\mathbb{H} =\sum_{j=0}^{r-1} b_j \mathbb{M}^j$, where the coefficients $b_j$ are to be determined. Substituting the latter in (\ref{A3}) and including (\ref{A2}), we obtain
$$ \sum_{j=1}^{r-1} \left(a b_j - \epsilon b_{j-1} + \epsilon b_{r-1} k_j\right) \,\mathbb{M}^j + (a b_0 +\epsilon b_{r-1} k_0)\,\mathbb{I} = \mathbb{I}  $$
which is equivalent to
$$ a b_j - \epsilon b_{j-1} + \epsilon b_{r-1} k_j =\delta^0_j \,, \qquad j=0 \ldots r-1
    \,, \quad b_{-1}=0  $$
To solve this system for $b_j$, $j=0,\ldots,r-1$, we write it in matrix form: 
\begin{equation} \label{A4} 
\mathbb{A} \vec b + \epsilon b_{r-1} \vec k = \vec v 
\end{equation} 
with $v_j=\delta^0_j\,$ and $\mathbb{A}=a\,\mathbb{I}-\epsilon\mathbb{N}$, where $\mathbb{N}=(N^i_j)\,$ with $N^i_j=\delta^{i+1}_j$.

Solving the above system  is easy once we realize that $\mathbb{N}^r=0$. Whence
$$\mathbb{A}^{-1} = \frac1a\,\sum_{j=0}^{r-1} \left(\frac{\epsilon}{a}\right)^j \,\mathbb{N}^j $$
and from (\ref{A4}) we have:
$$\vec b = \mathbb{A}^{-1} \vec v - \epsilon \,b_{r-1} \mathbb{A}^{-1}\vec k \,.$$
Now, $b_{r-1} = \vec w^{\rm T} \vec b$, with $w_j=\delta^{r-1}_j$, and it follows that
$$b_{r-1} \left(1 + \epsilon\, \vec w^{\rm T}\mathbb{A}^{-1}\vec k\right) = \vec w^{\rm T}\mathbb{A}^{-1}\vec v $$
which can be solved for $b_{r-1}$ whenever $1 + \epsilon\, \vec w^{\rm T} \mathbb{A}^{-1}\vec k \neq 0$. A more careful calculation that includes (\ref{A2}) shows that the latter is equivalent to
$$ \sum_{l=0}^{r-1} \left(\frac{\epsilon}{a}\right)^l \,k_l \neq 0 $$
i. e. $m(a\epsilon)\neq 0$, which holds whenever $\overline{g}_{\alpha\beta}$ is non-degenerate.

Below we list the explicit expressions for $\overline{h}^{\alpha\beta}$ in some few lowest dimensional cases ($n=3,4$):
\begin{eqnarray}  \label{A12}
\underline{n=4} &\qquad & \overline{h}^{\alpha\beta} = \frac{a+\epsilon I_1}{p(\epsilon a)} \,g^{\alpha\beta}+ \frac{\epsilon }{p(\epsilon a)}\, F^\alpha_{\;\mu} F^{\mu\beta}  \\  
& & \nonumber \\
\underline{n=3} &\qquad & \overline{h}^{\alpha\beta} = \frac1{a} \,g^{\alpha\beta}+ \frac{\epsilon }{p(\epsilon a)}\, F^\alpha_{\;\mu} F^{\mu\beta}   \nonumber
\end{eqnarray}
where $p(x)=x^2+I_1 x+I_2$.

The case $n=2$ is trivial because $M_{\alpha\beta}=I_1 \, g_{\alpha\beta}$ and $\overline{g}_{\alpha\beta}=(a+\epsilon I_1) \, g_{\alpha\beta}$.

\subsection*{Appendix B: The characteristic polynomials for $F^\alpha_{\;\beta}$ and $\overline{F}^\alpha_{\;\beta}$} 
We here restrict to the particular case $n=4$. The characteristic polynomials for $F^\alpha_{\;\beta}:=g^{\alpha\mu} F_{\mu\beta}$ and $\overline{F}^\alpha_{\;\beta}:= \overline{h}^{\alpha\mu} {F}_{\mu\beta}$ are, respectively,
$$ \chi_F(X)=X^4+I_1 X^2 + I_2 \qquad {\rm and} \qquad
\chi_{\overline{F}}(X)=X^4+\overline{I}_1 X^2 + \overline{I}_2 $$
In matrix notation, $\mathbb{F}=(F^\alpha_\beta)$ and $\overline{\mathbb{F}}= (\overline{F}^\alpha_\beta)$, we have 
\begin{equation} \label{A11} 
\mathbb{F}^4 +I_1  \mathbb{F}^2 +I_2 \mathbb{I} = 0
\qquad{\rm and} \qquad \overline{\mathbb{F}}^4+ \overline{I}_1
\overline{\mathbb{F}}^2+\overline{I}_2 \mathbb{I} = 0  \,.
\end{equation} 
$I_1$ and $I_2$ (resp., $\overline{I}_1$ and $\overline{I}_2$) are the $g$-invariant coefficients (resp., the $\overline{g}$-invariant coefficients) of the 2-form $F_{\alpha\beta}$.

Introducing now the matrix $\mathbb{H}=(\overline{h}^\alpha_\beta)$, with $\overline{h}^\alpha_\beta$ given by (\ref{A12}), we have
\begin{equation} \label{A13}
\overline{\mathbb{F}}=\mathbb{H}\mathbb{F} \qquad {\rm and} \qquad
\mathbb{H}=\frac{a+\epsilon I_1}{p(\epsilon a)}\,\mathbb{I}+\frac{\epsilon }{p(\epsilon a)}
\,\mathbb{F}^2
\end{equation}
Since $\mathbb{F}$ and $\mathbb{H}$ commute, the second equation (\ref{A11}) multiplied by ${\mathbb{H}}^{-4}=\overline{\mathbb{G}}^4$ yields: 
$${\mathbb{F}}^4 + \overline{I}_1 {\mathbb{F}}^2\overline{\mathbb{G}}^2+\overline{I}_2 \overline{\mathbb{G}}^4 = 0\,$$ 
which is a polynomial of eighth degree in ${\mathbb{F}}$ involving only even powers. The first equation (\ref{A11}) can then be used to reduce all powers of an order
higher than 2, so obtaining $C_1  \mathbb{F}^2 +C_0 \mathbb{I} = 0$, with:
$$ C_0=(a^2-I_2)\left(\overline{I}_2-\frac{I_2}{p(\epsilon a)}\right) \;, \qquad
C_1=\overline{I}_1-\frac{2a\epsilon+I_1}{p(\epsilon a)}\,\overline{I}_2-\frac{a^2I_1+2a\epsilon
I_2}{p(\epsilon a)^2}
$$
In order that the above matrix equation does not imply an extra restriction on $\mathbb{F}$ [besides the characteristic equation (\ref{A11})], the coefficients $C_0$ and $C_1$ must vanish identically. Therefore we have that:
\begin{equation} \label{A14}
\overline{I}_2=\frac{I_2}{p(\epsilon a)} \qquad {\rm and} \qquad
\overline{I}_1 = \frac{a^2 I_1+ 4 a \epsilon I_2 + I_1 I_2}{p(\epsilon a)^2} \,.
\end{equation} 

\subsection*{Appendix C: The characteristic determinant of the partial differential system (\ref{2.2b}-\ref{2.2d})}
We have to prove that Cauchy data $a$, $E_\alpha$ and $\tilde{F}_{\alpha\beta}$ can be provided such that the homogeneous linear system 
\begin{equation} \label{A22} 
\left. \begin{array}{l}
\ddot a \, \hat g_{\alpha\beta}  + \sigma\epsilon \,\left(\ddot E_\alpha E_\beta + E_\alpha \ddot E_\beta \right) - \epsilon\, \left(\ddot{\tilde{F}}_{\alpha\rho} \,\tilde{F}^{\rho}_{\;\beta} + \tilde{F}_{\alpha}^{\;\rho} \,\ddot{\tilde{F}}_{\rho\beta} \right) = 0 \\
\Psi_1\ddot a + \Psi_2^\alpha \ddot E_\alpha + \Psi_3^{\alpha\beta} \ddot{\tilde{F}}_{\alpha\beta} = 0
\end{array} \right\}
\end{equation} 
only admits the trivial solution, i. e. is determined. Although indices  $\alpha,\beta \ldots$ run through $1,\ldots,n$, the effective dimension of the linear system is $d=n-1$, as we are in the space tangential to ${\cal S}$, i.e. orthogonal to $n^\alpha$.

Assume that the Cauchy data $E_\alpha$ and ${\tilde{F}}_{\alpha\beta}$ are chosen so that the vectors:
$$ v_j^\alpha := (\tilde F^{j-1})^\alpha_{\; \beta} E^\beta \;, \quad j= 1,2, \ldots d$$
are linearly independent, so constituting a base for the vector space tangential to ${\cal S}$. Let $\omega^j_\alpha$, $j=1,\ldots d$, denote the dual base and  define:
\begin{eqnarray}
 & e_i:= E_\alpha \left(\tilde F^{i-1}\right)^\alpha_{\;\beta} E^\beta \;, &\quad
  G_{ij}:= \hat g_{\alpha\beta} v_i^\alpha v_j^\beta = (-1)^{i-1}e_{i+j-1} \;, \label{for1a} \\
& \xi_i:= \sigma\epsilon \ddot E_\alpha v_i^\alpha \;, &\quad
  Y_{ij}:= \epsilon \ddot{\tilde{F}}_{\alpha\beta} v_i^\alpha v_j^\beta = - Y_{ji}
\label{for1b} \\
 &\psi_2^i:=\sigma\epsilon \Psi_2^\alpha \,\omega^i_\alpha \;, & \quad
\psi_3^{ij}:=\epsilon \Psi_3^{\alpha\beta} \,\omega^i_\alpha \omega^j_\beta
\label{for1c} 
\end{eqnarray}
Since $\tilde{F}_{\alpha\beta}$ is skewsymmetric, $e_j=0\,$ for $j$ even. Moreover, the coefficients $e_j$ with $j>d$ depend on the coefficients $e_l\,$, $l\leq d$. Indeed, the characteristic polynomial, $\det\left(\tilde{F}^{\alpha}_{\;\beta}- X
\hat{g}^{\alpha}_{\;\beta}\right)$, allows to write:
\begin{equation} \label{rel1}
\tilde{\mathbb{F}}^d = \sum_{b=1}^D A_b \tilde{\mathbb{F}}^{d-2b}
\end{equation}
with $D=[d/2]$ and the coefficients $A_b$ are the invariant coefficients of $\tilde{F}^\alpha_{\; \beta}$  (as $\tilde{F}_{\alpha\beta}$ is skewsymmetric, (\ref{rel1}) only involves the powers $d-2b$). Therefore, for  $l>d$ we have the relations:
$$ e_l = \sum_{b=1}^D A_b\, e_{l-2b}  \qquad {\rm and} \qquad Y_{il}=\sum_{b=1}^D A_b\, Y_{i\;l-2b} $$
or, taking $A_0=-1$,
\begin{equation} \label{rel2}
\sum_{b=0}^D A_b\, e_{l-2b}=0 \qquad {\rm and}\qquad \sum_{b=0}^D A_b\, Y_{i\;l-2b}=0 \,, \qquad j>d
\end{equation}
which allows writing all coefficients $e_l$ and the variables $Y_{kl}$ in terms of $e_j$ and $Y_{ij}$, for $i,j\leq d$, and the characteristic polynomial coefficients $A_b$, $b= 1, \dots D$.

On contraction with $v_i^\alpha\, v_j^\beta$ the system (\ref{A22}) yields:
\begin{eqnarray} \label{A23a}
{\cal B}_{ij} &:=& \ddot a \, G_{ij}  + \xi_i e_j + \xi_j e_i + Y_{i+1,j} - Y_{i,j+1}  = 0 \\  \label{A23b}
{\cal B}_0 &:=& \Psi_1\ddot a + \sum^d_{i=1}\psi_2^i \xi_i + \sum^d_{i,j=1}\psi_3^{ij} Y_{ij} = 0
\end{eqnarray}

Since some among these equations are redundant, only those corresponding to $1\leq i\leq j \leq d$ must be considered. Note incidentally that, due to (\ref{rel2}), $Y_{i\,d+1}$ can be written in terms of $Y_{ij}$, with $i,j\leq d$. Therefore, (\ref{A23a}-\ref{A23b}) is a linear homogeneous system of $1+d(d+1)/2$ equations for the same number of unknowns (the coefficients depend on the scalars $e_l$, $A_b$, and on $\Psi_1$, $\psi_2^i$ and $\psi_3^{ij}$). 

On recombining the equations (\ref{A23a}-\ref{A23b}) we obtain the equivalent linear system:
\begin{eqnarray} \label{A24a}
{\cal B}_{ij} &=& 0  \qquad 1\leq i\leq j<d  \\  \label{A24b}
{\cal B}_{i} &:=& \sum_{l=0}^{D^\prime} A_l {\cal B}_{i\;d-2l} = 0 
\,,\qquad 1\leq i\leq d  \\  \label{A24c}
{\cal B}_0 &=&  0
\end{eqnarray}
with $D^\prime = D$ when $d\neq \dot{2}$ and $D^\prime = D-1$ when $d=\dot{2}$.

Equations (\ref{A24a}) can be solved in the $Y's$:
\begin{equation}  \label{A25}
Y_{il} = \ddot{a}\,\frac{(-1)^{i-1}}{2}\, e_{i+l-2} + \sum_{t=i}^{l-1}\xi_t e_{i+l-t-1} \,, \qquad i<l 
\end{equation}
which,substituted into equations (\ref{A24b}) and (\ref{A24c}), yields a linear homogeneous system in the unknowns $\ddot{a}$ and $\xi_j$, $j=1\ldots d$.

Now we have to proceed differently depending on whether $d$ is odd or even.

\paragraph{Case $d=1+2D$} On substituting (\ref{A25}) and using (\ref{rel2}), the system (\ref{A24b}-\ref{A24c}) becomes:
\begin{equation}  \label{A26}
\left. \begin{array}{lcl}
 {\cal B}_{2i}=0\,,\quad i=1\ldots D& \rightarrow & \displaystyle{\sum_{t=1}^D P_i^t \xi_{2t}=0} \\
 {\cal B}_{2i+1}=0\,,\quad i=1\ldots D& \rightarrow & \displaystyle{\sum_{t=1}^D (P_i^t -M_d \delta_D^t \delta_i^D) \xi_{1+2t}=0} \\
 {\cal B}_{1}=0\,, & \rightarrow & \displaystyle{\frac{M_d}{2}\,\ddot{a} + M_d \xi_1 + \sum_{t=2}^{2D+1} P_0^t\xi_t=0} \\
 {\cal B}_{0}=0\,, & \rightarrow & \displaystyle{\varphi_1\,\ddot{a} + \sum_{t=1}^{2D+1} \varphi_2^t \xi_t=0} 
       \end{array}  \right\}
\end{equation}
with
\begin{equation}  \label{A27a}
 M_d=\sum_{l=0}^D A_l e_{d-2l} \qquad {\rm and} \qquad
 P_i^t=\sum_{k=t}^D A_{D-k} e_{1+2(k+i-t)} \,, \qquad i,t=0\ldots D
\end{equation}
and 
\begin{equation}
\varphi_1 = \Psi_1 + \sum_{i<l} \psi_3^{il}\,(-1)^{i-1}\, e_{i+l-2} \quad {\rm and} \quad
\varphi_2^t= \psi_2^t + 2 \,\sum_{i=1}^t\sum_{l=t+1}^d \psi_3^{il} e_{i+l-t-1}
\end{equation}
It is now obvious that the system admits only the trivial solution if, and only if, its determinant does not vanish, i.e.
\begin{equation}  \label{A28}
\Delta(2D+1)=\frac{M_d}{2}\,\left(2\varphi_1 - \varphi_2^1 \right) \,\lambda(D)\,\rho(D)\neq 0 
\end{equation}
with $ \lambda(D):=\det\left(P_i^t\right)_{1\leq i,t\leq D}$ and  $\rho(D):=\det\left(P_i^t-M_d\delta_D^t \delta_i^D\right)_{1\leq i,t\leq D}$.

\paragraph{Case $d=2D$} Now, with $D^\prime = D-1$, the system (\ref{A24b}-\ref{A24c}) becomes:
\begin{equation}  \label{A29}
\left. \begin{array}{lcl}
 {\cal B}_{2i}=0\,,\quad i=1\ldots D & \rightarrow & \displaystyle{\sum_{t=1}^D P_i^t \xi_{2t-1}=0} \\
 {\cal B}_{2i-1}=0\,,\quad i=1\ldots D& \rightarrow & \displaystyle{\sum_{t=1}^D P_{i-1}^t \xi_{2t}=0} \\ 
 {\cal B}_{0}=0 \,, & \rightarrow & \displaystyle{\varphi_1\,\ddot{a} + \sum_{t=1}^{2D} \varphi_2^{t} \xi_t=0} 
       \end{array}  \right\}
\end{equation}
The system admits only the trivial solution if, and only if, its determinant does not vanish, i.e.
\begin{equation}  \label{A30}
\Delta(2D)=\varphi_1\,\lambda(D)\,\rho^\prime(D)\neq 0 
\end{equation}  
with $\rho^\prime(D):=\det\left(P_{i-1}^t\right)_{1\leq i,t\leq D}$.
 
\paragraph{Are $v_1^\alpha,\ldots,v_d^\alpha$ linearly independent?}
As far as the above reasoning relies on the assumption that the vectors $v_j^\alpha\,$, $j=1 \ldots d$ are linearly independent, the condition $\Delta(d)\neq 0$ must be supplemented with $\Gamma(d):=\det(G_{ij})\neq 0$. Now, since $G_{ij}=(-1)^{i-1} e_{i+j-1}= 0$ for $i+j\neq \dot 2$, this determinant results greatly simplified by rearranging rows and columns so that odd indices are written before even ones: 
$$ \Gamma(d) = 
   \left| \begin{array}{c|c}
         \left| G_{ij}\right|_{i,j\neq\dot 2} & 0 \\ \hline
         0 & \left| G_{ij}\right|_{i,j=\dot 2}
         \end{array}  \right|   $$
Therefore
$$ \Gamma(d) = \Gamma_{\rm odd}(d)\, \Gamma_{\rm even}(d) $$
with $\Gamma_{\rm even}(d)=\det\left( G_{2i\,2j}\right)_{i,j=1\ldots D}$ and $\Gamma_{\rm odd}(d)= \det\left( G_{2i+1\,2j+1}\right)_{i,j=1\ldots D^\prime}$, where $D^\prime =[(d-1)/2]$. 

Using then $G_{ij}=(-1)^{i-1} e_{i+j-1}$ and the fact that a determinant does not change on adding to each column a linear combination of the others, we readily obtain: $\,\Gamma_{\rm even}(d)=(-1)^{[D/2]}\,\lambda(D)$ and 
$$ \Gamma_{\rm odd}(d)= \left\{ \begin{array}{ll}
                             (-1)^{[D/2]+1}\,M_d\,\lambda(D) \,,& d=1+2D \\
                             (-1)^{[D/2]}\,\rho^\prime(D) \,,& d=2D
                             \end{array}  \right.   $$ 
Therefore
\begin{equation} \label{A31}
\Gamma(2D) = \lambda(D)\,\rho^\prime(D)    \qquad {\rm and} \qquad  \Gamma(2D+1) = -M_d\,\lambda(D)^2 
\end{equation}                    
Summarizing, a sufficient condition for the linear system (\ref{A22}) not to have non-trivial solutions is that:
\begin{equation}  \label{A31a}
\begin{array}{l|cccc}
  d\neq\dot 2 & M_d\neq 0\,,\; &  2 \varphi_1 -\varphi_2^1 \neq 0\,,\; &
                \lambda(D)\neq 0\,,\; &  \rho(D)\neq 0 \\ \hline
  d=\dot 2 &   & \varphi_1\neq 0 \,,\; &
                \lambda(D)\neq 0\,,\; &  \rho^\prime(D)\neq 0 
    \end{array}  
\end{equation}
Notice that the simplest possible choice for the constraint, namely, $\Psi(a,F):= a-1 =0$ is consistent with this condition for ${\cal S}$ to be non-characteristic. Indeed, for this constraint, $\psi_2^i=0$ and $\psi_3^{ij}=0$. Therefore, if $d\neq \dot 2$, $2\varphi_1-\varphi_2^1 = 2\neq 0$, and if $d= \dot 2$, $\varphi_1=1\neq 0$

Next, with an eventual application to general relativity ($n=4$) in view, we display some explicit expressions for conditions for the system (\ref{A22}) to be determined in some few lowest dimensional cases ($d=1,2,3$). We moreover assume that the constraint $\Psi$ is a scalar and therefore only depends on the invariants $I_1$, \ldots $I_D$ of $F^\alpha_{\;\beta}$.

\paragraph{\underline{$n=4$, $d=3$}, $D=1$,} Constraint $\quad \Psi(a,I_1,I_2)=0$. \\
In this case the condition (\ref{A31a}) reads
\begin{equation} \label{A32a}
A_1e_1-e_3 \neq 0 \,,\qquad e_3\neq 0 \,,\qquad A_1e_1 \neq 0
\end{equation}
and
\begin{equation} \label{A32b}
 2\varphi_1 -\varphi_2^1 = 2\,\left(\frac{\partial\Psi}{\partial a} - 2\epsilon \,\frac{\partial\Psi}{\partial I_1} + \epsilon A_1\, \frac{\partial\Psi}{\partial I_2} \right)\neq 0\,,
\end{equation}
where
$$
A_1 = \sigma e_1-I_1\,,\qquad e_3 = e_1 A_1 +\sigma I_2 \,,\qquad e_1 =-n_\mu F^\mu_{\;\alpha} F^\alpha_{\;\nu} n^\nu
$$
$$
I_1=- \frac12 \,F^{\alpha\beta} F_{\beta\alpha}\,, \qquad 
I_2=-\frac14 \,F^{\alpha\beta} F_{\beta\nu} F^{\nu\mu} F_{\mu\alpha} +\frac12 I_1^2 $$

\paragraph{\underline{$n=3$, $d=2$}} Constraint $\quad \Psi(a,I_1)=0$. \\
From (\ref{A31a}) we have that
\begin{equation} \label{A33a}
e_1 \neq 0 \,, \qquad A_1 \neq 0 \,, \qquad \varphi_1 = \frac{\partial\Psi}{\partial a} - \epsilon \,\frac{\partial\Psi}{\partial I_1} \neq 0
\end{equation}
where
$$
A_1 = \sigma e_1-I_1\,,\qquad {\rm and}\qquad 
I_1=- \frac12 \,F^{\alpha\beta} F_{\beta\alpha} $$

\paragraph{\underline{$n=2$, $d=1$}} Constraint $\quad \Psi(a,I_1)=0$. \\
In this simple case the condition can be directly obtained from (\ref{A22}) and reads:
\begin{equation}  \label{A34a}
e_1 \neq 0 \,, \qquad \varphi_1 = \frac{\partial\Psi}{\partial a} - \epsilon \,\frac{\partial\Psi}{\partial I_1} \neq 0
\end{equation}


\section*{References}
\begin{thebibliography}{99}
\bibitem{Riem}
G.~F.~B. RIEMANN., ``On the hypotheses which lie at the foundations of geometry'', in D.~E. SMITH, editor, {\em A source book in mathematics}. Dover, (New York, 1959)

\bibitem{Eissen}
L.~P. EISENHART., {\em A Treatise on the Differential Geometry of Curves and Surfaces},
Dover, (New York, 1960)

\bibitem{CllS01}
J.~COLL, B.;~LLOSA and D.~SOLER., {\em Gen. Rel. and Grav.}, {\bf 34} (2002) 269

\bibitem{Diag1}
 For an analytic proof on the existence of orthogonal coordinates, see E.~CARTAN., {\em Les syst\`emes differentiels ext\'erieurs et leurs applications g\'eom\'etriques}, Hermann, (Paris, 1971)

\bibitem{Diag2}
M.~DETURCK and D.~YANG., {\em Duke Math. Jour.}, {\bf 51} (1984) 243

\bibitem{Diag3}
P.~WALBERER., {\em Abhandl. Math.Sem. Univ. Hamburg}, {\bf 10} (1934) 169

\bibitem{Diag4}
L.~BEL., {\em Gen. Rel. and Grav.}, {\bf 28} (1996) 1139

\bibitem{Diag5}
S.~TANNO., {\em J. Differential Geometry}, {\bf 11} (1976) 467

\bibitem{tolo}
B.~COLL., ``A universal law of gravitational deformation for general relativity'', in J.~Martin, E.~Ruiz, F.~Atrio, and A.~Molina, editors, {\em Gravitation and Relativity in General}, World Scientific, (1999).

\bibitem{Weinberg}
S.~WEINBERG., {\em  Gravitation and Cosmology: Principles and aplications of the
  General Relativity Theory}, John Willey \& Sons, (New York, 1972)

\bibitem{John}
F.~JOHN., {\em Partial Differential Equations}, Chap. 3, Springer, (New York, 1971).

\bibitem{Hicks}
N.~J. HICKS., {\em Notes on differential geometry}, Van Nostrand, (Princeton, 1965)

\bibitem{Choquet}
Y.~CHOQUET-BRUHAT., {\em Analysis, Manifolds and Physics}, (North-Holland, 1980).

\bibitem{KerrSchild}
R.~P. KERR and A.~SCHILD., {\em J. Math. Phys.}, {\bf 10} (1969) 1842

\bibitem{SenoHilde}
HILDEBRAND S.~R. COLL, B. and J.~M.~M. SENOVILLA, {\em Gen. Rel. and Grav.}, {\bf 33} (2001) 649

\bibitem{NashMoser}
R.~HAMILTON., {\em Bulletin Amer. Math. Soc. }, {\bf 7} (1982) 65

\end{thebibliography}
\end{document}